\documentclass[reprint,amsmath,amssymb,doublecolumn,aps,prb]{revtex4-1}
\usepackage{graphicx}
\usepackage{dcolumn}
\usepackage{bm}
\usepackage{makecell}
\usepackage{tabularx}
\usepackage[utf8]{inputenc}
\usepackage[english]{babel}
\usepackage{color,soul}
\usepackage{gensymb}
\begin{document}
\preprint{APS/123-QED}
\title{Zener Pinning through Coherent Precipitates: A Phase Field Study}
\author{Tamoghna Chakrabarti$^1$}%
\email{tamoghna@iitp.ac.in}
\author{Sukriti Manna$^{2}$}
\affiliation{$^1$Dept. of Materials Science and Engineering, Indian Institute of Technology Patna, Patna, 801106, India\\
$^2$Dept. of Mechanical Engineering, Colorado School of Mines, Golden, Colorado 80401, USA\\}

\date{\today}
\begin{abstract}
	A novel phase field model has been developed to study the effect of coherent precipitate on the Zener pinning of matrix grain boundaries. The model accounts for misfit strain between precipitate and matrix as well as the elastic inhomogeneity and anisotropy between them. The results show that increase in elastic misfit, elastic inhomogeneity, and elastic anisotropy increases the coarsening rate of the precipitates. Increased coarsening of precipitates in turn decreases the pinning of grain boundaries. Therefore, increase in misfit strain, elastic inhomogeneity and anisotropy mostly negatively affect the Zener pinning through coherent precipitate. This study shows elastic anisotropy gives rise to the needle shape precipitate. It has also been shown that these needle shaped precipitates are not very effective in Zener pinning. This study provides an understanding into the effect of coherent precipitate on the Zener pinning of matrix grain boundaries. To design a material with smallest possible grain size, coherent precipitate with least lattice misfit and highest elastic modulus will be most effective.
\end{abstract}
\maketitle
\section{\label{sec:level1}Introduction}
The grain size and morphology in polycrystalline materials often play an important role in determining the properties of materials. A wide gamut of material properties such as yield strength\cite{lasalmonie1986influence} and ultimate tensile strength,\cite{hansen1982strain} creep,\cite{sherby2002influence} fracture resistance,\cite{becher1998microstructural} oxidation,\cite{raman1992influence} corrosion resistance,\cite{argade2012effects,ralston2010revealing} electrical,\cite{tschope2001grain} magnetic,\cite{xuan2009tuning} and optical,\cite{okazaki1973effects} properties can be altered by modifying the grain size and morphology. In systems such as aluminum,\cite{JONES1981589,TWEED19841407} aluminum based alloys,\cite{SKChang} micro-alloyed steels \cite{guo1999pinning}, and nickel based superalloys\cite{MURALIDHARAN1997755}, reduction in grain size improves the mechanical strength. Thus, it is imperative to understand the external factors such as solute segregation, precipitate nucleation or second phase addition by which we can control grain sizes and morphology to obtain desired properties depending on applications. 

One of the simplest and possibly the most profound way of controlling the grain size is by means of introducing second phase in form of particles or precipitates.\cite{hillert1988inhibition} The role of second phase particles or precipitates on the refining of the grain size has been first proposed by Smith and Zener\cite{smith1948grains,pa1998five}. This effect which is called Zener pinning works by pinning the grain boundaries at the particle-grain interface. In Zener pinning, second phase size, morphology, volume fraction, coherency, anisotropic interfacial energy and even the coarsening of the second phase particle affect the final pinning.\cite{Huang2016} 

Theoretical models have been extensively employed to understand the interactions of these afore mentioned parameters in Zener pinning. Different computational techniques such as Monte-Carlo Potts models,\cite{soucail1999monte} front-tracking-type models,\cite{vanherpe2010pinning} and phase field models\cite{moelans2006phase,suwa2006phase} has been used to understand the Zener pinning phenomenon. Among these simulation techniques, we concentrate on phase field model to understand the Zener pinning phenomenon. The phase field model is a diffuse-interface model where the evolution of arbitrary complex grain and precipitate morphologies can be studied without any presupposition on their shape or distribution. Additionally, phase field model simulation results have also shown to be qualitatively consistent with experimental observations in many different types of systems and problems. \cite{chakrabarti2017grain,mukherjee2011thermal,Mukherjee2016,tang2016formation,ghosh2017particles} 

Previously through phase field model, the effect of volume fraction, shape, size, anisotropy and coarsening of the second phase particles on Zener pinning has already been studied in detail.\cite{vanherpe2010pinning,chang2009effect,suwa2006phase} One of the important aspect to consider is the coherency of the second phase particles in Zener pining which has not been studied in much detail. Due to coherency between the precipitate and matrix, the misfit strain can induces elastic stress, which can alter the coarsening behavior of the precipitates. Changes in the coarsening kinetics of the precipitate in turn can influence the Zener pinning drag. 

Wang \textit{et. al.}\cite{wang2015phase} have investigated through a phase field model the effect of misfit strain of the Zener pinning by coherent precipitates. In that model, the inhomogeneity in modulus between the matrix and second phase particles was not included. But, in practical applications, the precipitate modulus can be significantly different from that of the matrix phase. Such inhomogeneity in elastic modulus can in turn change the coarsening kinetics which has been implemented in our model. Additionally, we have inspected the effect of anisotropy in the elastic modulus which has a significant influence in changing the precipitate morphology from globular to needle shape. 

Our article is organized as follows: In the section \ref{sec:level2}, we present the details our phase field model. In the section \ref{sec:level3}, we have discussed the results of our phase field simulations by systematically investigating the effect of of misfit strain, elastic inhomogeneity, and anisotropy on grain coarsening kinetics. Additionally, we have studied the effect of different particle morphology which arises from the elastic anisotropy. Finally, the section \ref{sec:level4} contains the succinct conclusions of our work.

\section{\label{sec:level2}Theoretical Formulations}
In our phase field model, the microstructure consists of two phases \textit{i.e.} matrix and second phase particles (or precipitates). There exists a misfit between second phase and matrix. The matrix is polycrystalline whereas precipitate is single crystalline.

\subsection{Free Energy Functionals}
 The total free energy ($F_t$) of system described by the sum of chemical energy ($F_{ch}$) and elastic energy ($F_{el}$) \textit{i.e.}
$F_{t} = F_{ch} + F_{el}$. $F_{chem}$ of the system with inhomogeneities in the field of $c(\mathbf{r},t)$ and $\eta_{i} (\mathbf{r}, t)$; $i = 1, 2, ... , n$ describes $n$ unique grain orientations in the matrix phase is given by,
\begin{equation}
F_{ch}=N_{v}\int_{v}[f_{0}(c,\eta_{i})+\kappa_c(\nabla c)^{2} +\sum_{i=1}^{n} \kappa_{i}^{\eta}(\nabla \eta_{i})^{2}]dv
\end{equation}
Where, $N_{v}$: number of molecules per unit volume, $f_{0}(c, \eta_{i})$: bulk free energy density,
$\kappa_{c}$: gradient energy co-efficient due to composition $c(\mathbf{r}, t)$ variable, $\kappa_{i}^{\eta}$: gradient energy coefficient due to order parameter $\eta_{i}(\mathbf{r}, t)$ variable, $v$: volume of our domain of interest, $\mathbf{r}$: real space vector. The bulk free energy density $f_{0}(c, \eta_i)$ is given by,\cite{mukherjee2011thermal}
\begin{equation}
f_{0}(c,\eta_i) = Ac^2(1-c)^2 +Bc^2\zeta(\eta_i)+Z(1-c)^2\sum_{i=1}^{n}\eta_{i}^{2}
\label{eq:e102}
\end{equation}
Where, $\zeta(\eta_i)$ is expressed as\cite{mukherjee2011thermal},
\begin{equation}
\zeta(\eta_i) = \sum_{i=1}^{n}[\frac{\eta_i^4}{4}-\frac{\eta_i^2}{2}+2\sum_{j>i}^{n}\eta_i^2\eta_j^2]+0.25
\label{e3}
\end{equation}

and the parameters $A, B, Z$ in equation (\ref{eq:e102}) are constants.  

\subsection{Elastic Energy}
In our phase field model, the source of misfit arises due to the compositional heterogeneity between the matrix and precipitate. This misfit introduces the elastic energy in the system. The elastic energy contribution of the total free energy is given by,
\begin{equation}
F_{el}=\frac{1}{2}\int_{v}\sigma^{el}_{ij}(\mathbf{r})\epsilon^{el}_{ij} (\mathbf{r})dv
\end{equation}
Where, $\sigma_{ij}^{el}(\mathbf{r})$: elastic stress tensor, $\epsilon^{el}_{ij}(\mathbf{r})$: elastic strain tensor and $\epsilon^{el}_{ij}(\mathbf{r})$ given by following equation,

\begin{equation}
\epsilon^{el}_{ij}=\epsilon_{ij}(\mathbf{r})-\epsilon_{ij}^{0}(\mathbf{r})
\end{equation}
Where, $\epsilon_{ij}^{0}(\mathbf{r})$: position dependent eigenstrain, $\epsilon_{ij}(\mathbf{r})$: total strain which is given by following equation,
\begin{equation}
\epsilon_{ij}(\mathbf{r})=\frac{1}{2}\left(\frac{\partial u_{i}}{\partial r_{j}}+\frac{\partial u_{j}}{\partial r_{i}}\right)
\end{equation}
Assuming that, there is no rotational component to the displacement field and phases obey the Hooke’s law (\textit{i.e.} both phases are linear elastic). Hence,
\begin{equation}
\sigma_{ij}^{el}(\mathbf{r})=C_{ijkl}(\mathbf{r})\epsilon^{el}_{kl} (\mathbf{r})
\label{e10}
\end{equation}
Where, $C_{ijkl}(\mathbf{r})$: elastic modulus tensor.
Now substituting the values of $\epsilon_{ij}^{el}(\mathbf{r})$ from equation (\ref{e10}) we obtain,
\begin{equation}
\sigma_{ij}^{el}(\mathbf{r})=C_{ijkl}(\mathbf{r})\left\lbrace\epsilon_{kl}(\mathbf{r})-\epsilon_{kl}^{0}(\mathbf{r})\right\rbrace
\end{equation}
As the stress field obeys the equation of mechanical equilibrium. Hence,
\begin{eqnarray}
\sigma_{ij,j}^{el}(\mathbf{r})=0 \nonumber \\
i.e.~~\frac{\partial}{\partial r_{j}}\left[ C_{ijkl}(\mathbf{r})\left\lbrace\epsilon_{kl}(\mathbf{r})-\epsilon_{kl}^{0}(\mathbf{r})\right\rbrace\right]=0
\end{eqnarray}
eigen strain, $\epsilon_{ij}^{0}(\mathbf{r})$ expressed as,
\begin{equation}
\epsilon_{ij}^{0}(\mathbf{r})=\theta^{c}(c)\epsilon^{c}\delta_{ij} 
\end{equation}
here $\theta^{c}(c)$ is a shape function which is approximated as a linear function (following Vegard's law\cite{denton1991vegard}) and expressed as, 
\begin{eqnarray}
\theta^{c}(c)=\frac{c(\mathbf{r})- c^{ppt}}{c^{ppt}-c^{mat}} 
\end{eqnarray}
$\theta^{c}(c)$  give value $1.0 $ at precipitate and $0.0$ at matrix. In between particle and matrix (interface region), it takes values in between $1.0$ and $0.0$, $\epsilon^{c}$: misfit strain between precipitate and matrix, $\delta_{ij}$: Kronecker delta.

We are solving the equations for a plane strain approximation \textit{i.e.} there is no eigenstrain in the $z$-direction.

\begin{table}[!htbp]
\centering
\caption{Simulation Parameters}
\label{table:simulation param}
\begin{tabularx}{0.8\columnwidth}{c c}
\hline \hline
\makecell{Simulation parameters}   & Values  \\
\hline
\makecell{Grid spacing in \\ $x$-direction ($\bigtriangleup x$)} & 0.5 \\
\hline
\makecell{Grid spacing in \\ $y$-direction ($\bigtriangleup y$)} & 0.5 \\
\hline
Timestep ($\bigtriangleup t$)  & 0.2 \\
\hline 
\makecell{System size\\in $x$-direction}  & 1024\\
\hline
\makecell{System size\\in $y$-direction}  & 1024\\
 \hline
 Mobility($M$)  & 1.0\\
 \hline
 \makecell{Relaxation \\ co-efficient($L$)}  & 1.0\\
 \hline
 $\kappa_{c}$, $\kappa_{\eta}$  & 1.0\\
 \hline
 \makecell{Shear modulus\\of matrix phase\\$(G^{mat}/N_{v})$}  & 400\\
 \hline
 Poisson's ratio & 0.3\\
 \hline
 misfit strain ($\epsilon$)  & $0.125\%-1.0\%$\\
 \hline
 $\delta=G^{ppt}/G^{mat}$  & 0.5-1.5\\
 \hline
 \makecell{Precipitate \\ initial radius}  & 4.0\\
 \hline
 $A, B, Z$ in equation.(\ref{eq:e102}) & 1.0 \\
\hline \hline
\end{tabularx}
\end{table}

\subsection{Kinetics of Microstructure Evolution}

We numerically solve Cahn-Hilliard equation for the evolution of composition  $c(\mathbf{r}, t)$ and Allen-Cahn equation for order parameters $\eta_{i}(\mathbf{r}, t)$.

Cahn-Hilliard equation\cite{cahn1961spinodal} is given by,
\begin{equation}
\frac{\partial c}{\partial t}= \nabla\cdot M\nabla \mu
\end{equation}
$M$ is mobility and is not a function of composition $(c)$ and order parameter $(\eta)$, $\mu$ is chemical potential and defined as, 
\begin{equation}
\mu=\frac{\delta}{\delta c}(F/N_{v})=\frac{\delta}{\delta c}(F_{ch}/N_{v})+\frac{\delta}{\delta c}(F_{el}/N_{v})
\end{equation}
Here, $\frac{\delta}{\delta c}$ represents the variational derivative with respective to composition.
\begin{equation}
\frac{\delta}{\delta c}(F_{ch}/N_{v})=\frac{\partial f_{0}}{\partial c}-2\kappa_{c}\nabla^{2}c
\end{equation}
The final form of Cahn-Hilliard equation will be,
\begin{equation}
\frac{\partial c(\mathbf{r},t)}{\partial t}=M\nabla^{2}\left(\frac{\partial f_{0}}{\partial c}-2\kappa_{c}\nabla^{2}c +\frac{\delta}{\delta c}(F_{el}/N_{v})\right)
\label{eq11}
\end{equation}

We use Allen-Cahn equation\cite{allen1979microscopic} for the evolution of the order parameters,
\begin{equation}
\frac{\partial \eta_{i}}{\partial t}=-L\frac{\delta}{\delta \eta_{i}}(F/N_{v})
\end{equation}
Where, $L$ is relaxation coefficient. $\frac{\delta}{\delta \eta_{i}}$ represents the variational derivative with respective to $\eta_{i}$.
Now,
\begin{equation}
\frac{\delta}{\delta \eta_{i}}(F/N_{v})=\frac{\delta}{\delta \eta_{i}}(F_{ch}/N_{v})+\frac{\delta}{\delta \eta_{i}}(F_{el}/N_{v})
\end{equation}
and 
\begin{equation}
\frac{\delta}{\delta \eta_{i}}(F_{ch}/N_{v})=\frac{\partial f_{0}}{\partial \eta_{i}}-2\kappa^{\eta}_{i}\nabla^{2}\eta_{i}
\end{equation}
The final form of Allen-Cahn equation is
\begin{equation}
\frac{\partial \eta_{i}(\mathbf{r},t)}{\partial t}=-L\left(\frac{\partial f_{0}}{\partial \eta_{i}}-2\kappa^{\eta}_{i}\nabla^{2}\eta_{i} +\frac{\delta}{\delta \eta_{i}}(F_{el}/N_{v})\right)
\label{eq12}
\end{equation}

Periodic boundary condition and semi-implicit Fourier spectral method were used to solve the PDEs. Discrete Fourier transformations were carried out using FFTW package.\cite{frigo2005design} Parameters used in the simulations are shown in Table \ref{table:simulation param}. We have used Hoshen-Kopelman algorithm\cite{hoshen1976percolation} to calculate the area of all the matrix grains and second phase precipitates.


\section{\label{sec:level3} Results and Discussions}
\begin{figure}[htp]
\begin{center}
\includegraphics[width=\columnwidth]{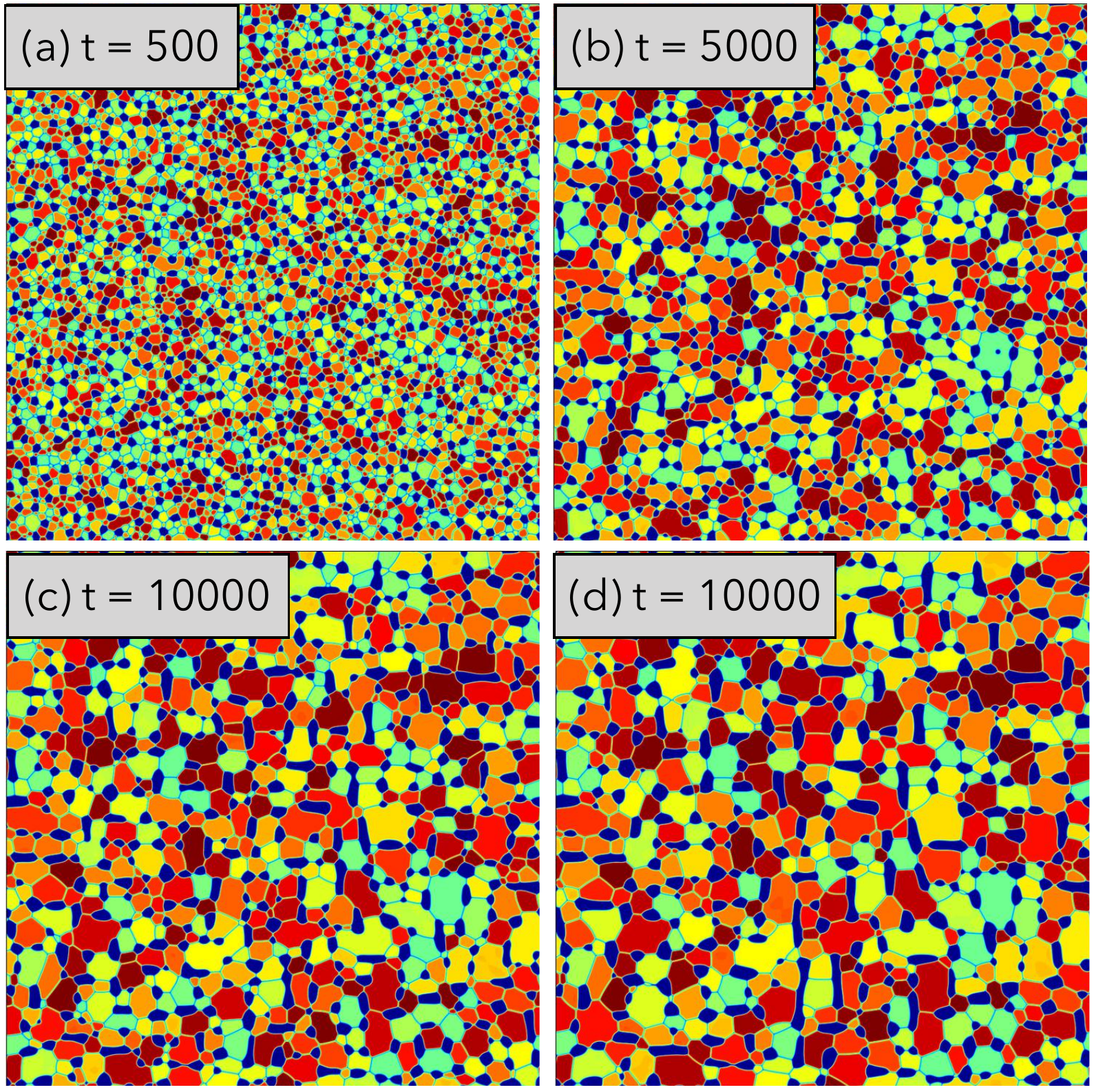}
\caption{Evolution of microstructures with  $\delta$ = 0.5 and $A_z$  = 4.0 at time step (a) $t =$ 500, (b) $t =$ 5000, (c) $t =$ 10000, and (d) $t =$ 14000 with second phase coherent particles with the misfit strength of $\epsilon=1.5\%$.  The parameters $\delta$ and $A_z$ denote the degree of elastic inhomogeneity and elastic anisotropic parameter respectively. The coherent particles are shown in blue color.\label{timeevolution}}
\end{center}
\end{figure}

In this work, we have extended the phase field model of Wang \textit{et. al.}\cite{wang2015phase} by incorporating inhomogeneity and anisotropy (cubic) in elastic modulus between grain and second phase precipitate to investigate  their effect on grain coarsening. The degree of elastic inhomogeneity and anisotropy are expressed by the parameters $\delta$ and $A_z$ respectively. The simulated micro-structures of grain coarsening in presence of coherent second phase precipitates are described in Fig.~\ref{timeevolution}(a)-(d). Here, we have used $\delta$=0.5 and $A_z$=4.0 to simulate the microstructures of Fig.~\ref{timeevolution}. Initially, ($t$=500), the second phase precipitates are randomly distributed on the matrix. With increasing time (Fig.~\ref{timeevolution} (a)-(d)), the particles starts aligning in $<10>$ directions. In the later stage ($t$=14000), the alignment of second phase precipitates are more prominent.  The second phase precipitates are gradually changing to rod like shapes from their initials globular shapes. Due to Ostwald ripening,\cite{thornton2004large} less numbers of second phase precipitates are observed in the later stages microstructures compared to its early stages. The size of grains increases with time and changes their morphology from globular to polygonal shape.

The rest of the results and discussions part is organized as follows: In section ~\ref{misfitsec}, the effect of misfit strain on grain growth kinetics is discussed. Next, in sections ~\ref{deltasec} and \ref{azsec}, we have investigated the effect of elastic inhomogeneity and anisotropy on grain growth kinetics. Finally, we end this section by discussing the combined effect (section~\ref{combinesec}) of elastic inhomogeneity and anisotropy on grain growth kinetics.   

\subsection{Effect of misfit strain ($\epsilon$)}\label{misfitsec}
\begin{figure}[htp]
\begin{center}
\includegraphics[width=\columnwidth]{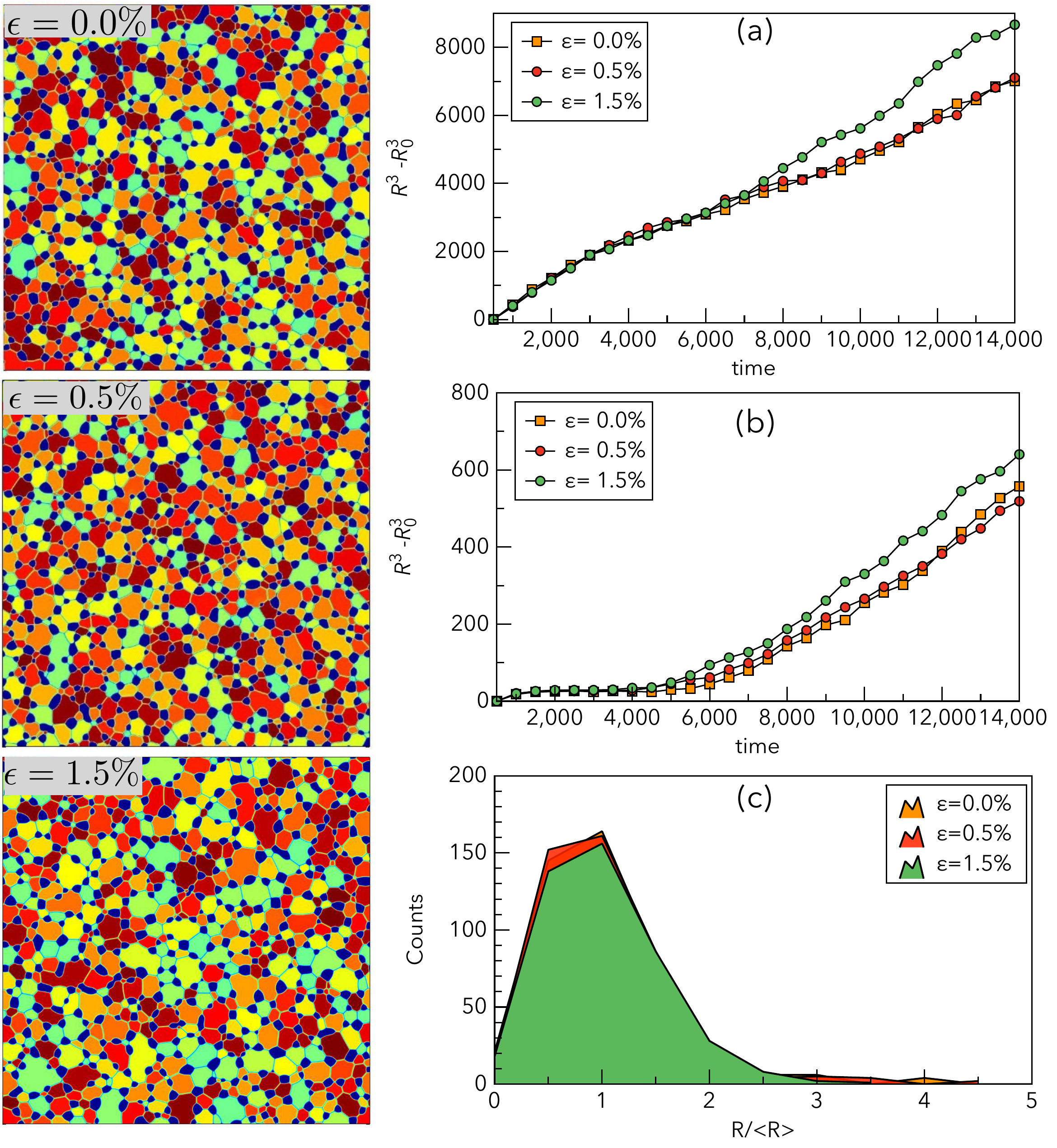}
\caption{Effect of misfit strain ($\epsilon$) on grain coarsening. Left column describes the final microstructures with different strength of $\epsilon$. Right column describes the effect of $\epsilon$ on (a) the coarsening kinetics of grains, (b) the coarsening kinetics of precipitates, and (c) the histogram of grain size distribution.\label{misfit}}
\end{center}
\end{figure}

Here, we have exclusively investigated the effect of misfit strain on microstructural evolution and grain coarsening kinetics. The final evolved microstructures with varying misfit strengths are shown in the left panel of Fig.~\ref{misfit}. Qualitatively, the grains and second phase particles in each microstructure with three different misfit strains are showing similar morphological characteristics. We have also calculated the average grain size and second phase precipitate size with three different misfit strains using Hoshen-Kopelman algorithm.\cite{hoshen1976percolation} The time evolution of average grain and precipitates sizes are described in Fig.~\ref{misfit} (a) and (b) respectively.
 
In Fig.~\ref{misfit} (b), we observe that the precipitate sizes remain similar till $t=$3000 in all three different misfit strain and and beyond $t=$ 3000 they diverge. In a system with coherent interface, there are two important contributions to the total energy namely elastic energy and chemical interfacial energy. As the precipitates grow, the elastic energy increases as $r^3$ and the chemical interfacial energy increases as $r^2$ with increasing precipitate size\cite{lund2002effects} where $r$ is the radius of the precipitate. Gradually, with continues growth of the precipitate, the elastic energy progressively dominates coarsening process. At initial time period (till $t=$3000) the coarsening process is controlled by chemical interfacial energy and resultantly in all the cases regardless of the misfit strain, precipitates coarsen at similar rate as system with no misfit ($\epsilon^c=0.0\%$). With time, as the elastic energy starts to become dominant, the coarsening rate of the coherent precipitates diverges. We also observe that at a later stage i.e. when elastic energy is dominating, the $\overline{r^3}$ ($\overline{r}$ is the average radius of the precipitate) shows linear relationship with time. Interestingly, such relationship has also been predicted by Larala \textit{et.al.}\cite{larala1989kinetics} with their analytical model.

From  Fig.~\ref{misfit} (b), it is evident that in presence of larger misfit, the coarsening of the precipitate gets faster compared to the system with no misfit strain. The increase in misfit strain changes the interfacial energy and also the equilibrium compositions of the phases. Thus increasing the misfit strain can influence the coarsening kinetics.\cite{larala1989kinetics} Here, we observe that the increase in misfit strain enhances the coarsening kinetics of the precipitates. Similar observations have also been made by Wang \textit{et.al.}.\cite{wang2015phase} They have shown through a phase field model that the elastic strain energy affects the solute diffusion.

Role of particle size on the matrix grain size can be established from the well known Zener formulation\cite{Huang2016} which is shown in equation (\ref{zener}). 

\begin{equation}
\overline{R_{lim}}  = \frac{4 r \alpha} {3 V_f }
\label{zener}
\end{equation}

Here, $r$ represents precipitate radius, $V_f$ is the volume fraction of second phase precipitates, and  the precipitate shape factor is expressed by $\alpha$. From this equation (\ref{zener}), we see that the systems with larger precipitate is less effective in pinning the grain. From our results we observe that the systems with larger misfit give rise to larger precipitate (as shown in the  Fig.~\ref{misfit} (a)). Due to lesser pinning by larger precipitate, the matrix grain size is also larger.

Fig.~\ref{misfit} (c) shows the final matrix grain size distribution. In all three cases with different misfit strain, the nature of matrix grain size distributions plots are similar. In the system with larger misfit, as the average grain size of matrix is larger, correspondingly number of grains are smaller. Therefore, peak height of the distribution plot is shorter in a system with large misfit strain compared to system with no misfit strain.

\subsection{Effect of elastic inhomogeneity ($\delta$)}\label{deltasec}
\begin{figure}[htp]
\begin{center}
\includegraphics[width=\linewidth]{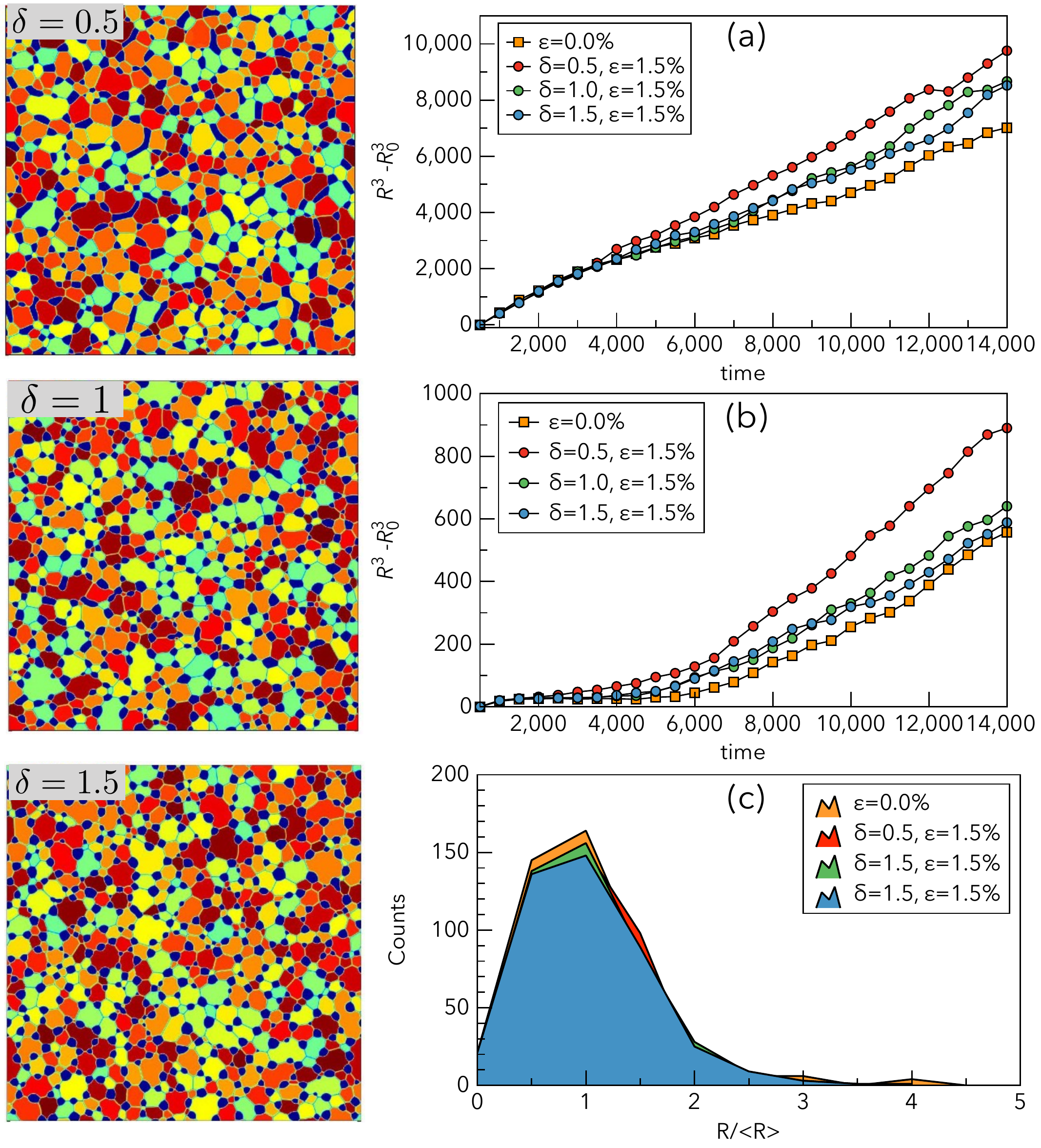}
\caption{Effect of elastic inhomogeneity ($\delta$) on the grain coarsening kinetics. Left column describes the final micro-structures with different degree of elastic inhomogeneity. On the right column, (a) the coarsening kinetics of grain with systems with different $\delta$ values, (b) coarsening kinetics of precipitates with different $\delta$ values, and (c) histogram of grain size distribution\label{inhomogeneity}}
\end{center}
\end{figure}
In this section, we have exclusively investigated the effect of elastic inhomogeneity ($\delta$) on grain coarsening kinetics by keeping other parameters (strength of misfit, elastic anisotropy parameter) constant. $\delta$=1.0 represents that the matrix phase and the precipitates are elastically homogeneous. If $\delta>$1.0, the precipitates are stiffer than matrix phase and  $\delta<$1.0 represents the precipitates which are softer than matrix phase. The final evolved micro-structures with different $\delta$ are shown in the left panel of the Fig.~\ref{inhomogeneity}. We observe that, in case of matrix stiffer than precipitate ($\delta$=0.5), the precipitates are becoming elongated. Here the stiffer matrix phase is likely exerting force on the more pliant precipitates and elongating them to minimize the total interfacial energy of the system. Such elongations are not observed in the other two cases ($\delta$=1.5 and 1.0) as the precipitates are stiffer or equal in modulus to the matrix phase. Time vs average matrix grain size and precipitate size data are shown Fig.~\ref{inhomogeneity} (a) and (b) respectively. 

Fig.~\ref{inhomogeneity} (a) shows largest average grain size in case of ($\delta$=0.5) and smallest in case of ($\delta$=1.5). With increasing value of $\delta$, the size of precipitate decrease. As described in the previous section, the precipitate size dictates the matrix grain size \textit{i.e.} the effectiveness of Zener pinning. Our study implies that the stiffer particles are more effective in Zener pinning than the softer one. 

\subsection{Effect of anisotropy in elasticity ($A_Z$)}\label{azsec}
\begin{figure}[htp]
\begin{center}
\includegraphics[width=\linewidth]{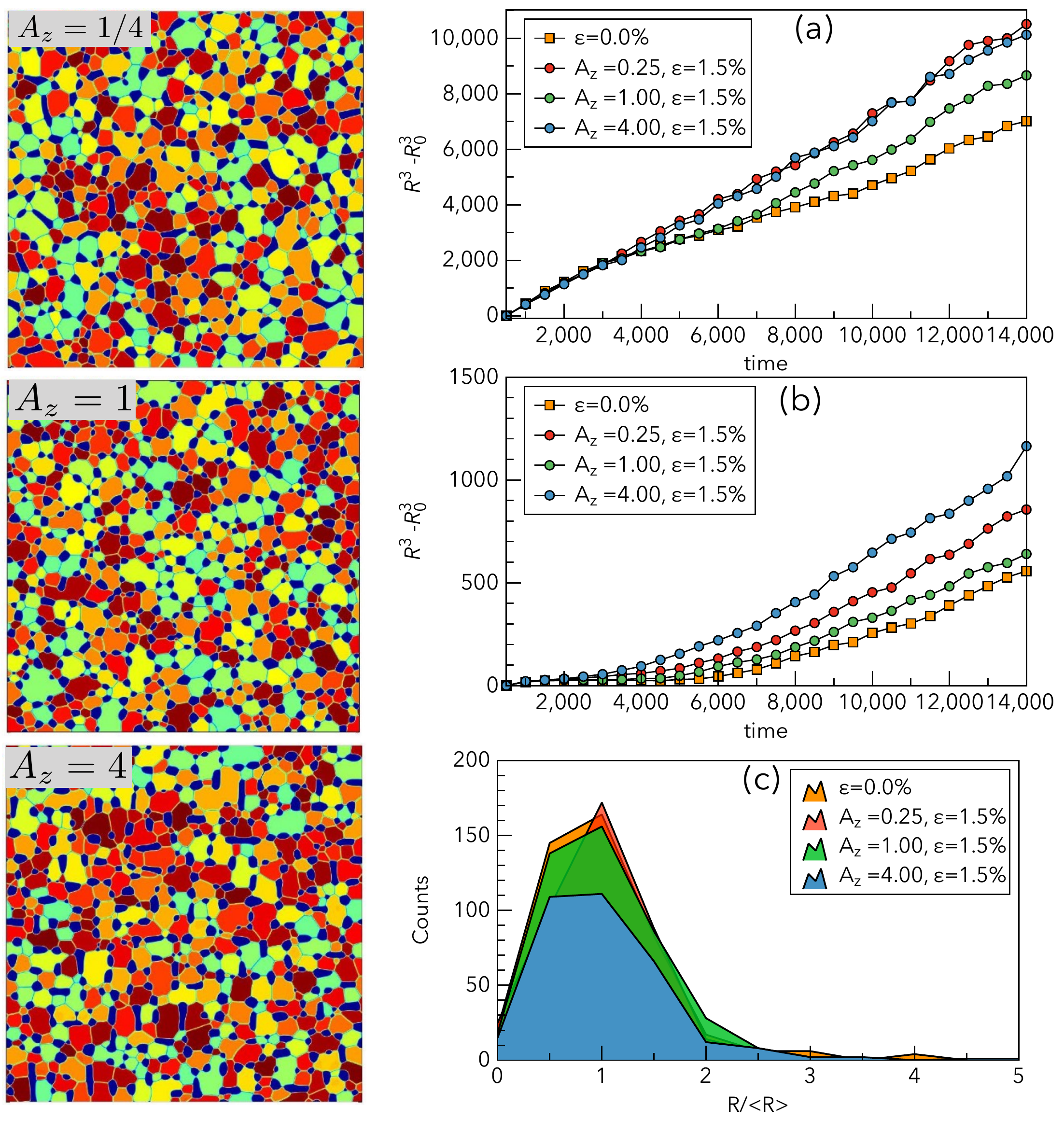}
\caption{Effect of elastic anisotropy parameter ($A_z$) on the grain coarsening kinetics with constant $\delta=1$ and constant misfit ($\epsilon=1.5\%$). Left column describes the final microstructures with several different anisotropy parameter, $A_z$. On the right column effect of $A_z$ on, (a) the coarsening kinetics of grain, (b)the coarsening kinetics of precipitate, and (c) the histogram of grain size distribution\label{anisotropy}}
\end{center}
\end{figure}
In addition to the study of the effect of misfit strain and elastic inhomogeneity (between grains and precipitates), we also have investigated the effect of elastic anisotropy on grain coarsening kinetics. In our model system, the elastic anisotropy is expressed by the Zener anisotropy parameter\cite{li1987single}($A_z$). The three independent elastic constants $(C_{11}, C_{12}$, and $C_{44})$ in a cubic anisotropic system can be related to the shear modulus $G$, the Poisson's ratio $\nu$, and the anisotropy parameter $A_z$ following equations \ref{eq:nu},\ref{eq:G}, and \ref{eq:az}.
\begin{equation}
\nu  = \frac{1}{2}\frac{2 \overline{c_{12}}} {\overline{c_{12}}+\overline{c_{44}}}
\label{eq:nu}
\end{equation}
\begin{equation}
G = \overline{c_{44}}
\label{eq:G}
\end{equation}
\begin{equation}
A_z=\frac{2 \overline{c_{44}}} {\overline{c_{11}}-\overline{c_{12}}}
\label{eq:az} 
\end{equation}

If $A_z$=1, then the system is considered to be elastically isotropic. $A_z>1$ means the $<10>$ direction is the softest direction. On the other hand, if $A_z<1$, then the $<11>$ direction is the elastically softest direction. In Fig.~\ref{anisotropy}, we have shown the effect of elastic anisotropy on the microstructural evolution. In the isotropic case i.e. $A_z$=1, we see that the particles are arranged in the matrix randomly whereas in case of $A_z$=1/4, a small level of preferential alignment of particles in $<11>$ direction are observed. On the other hand, we observe a relatively higher level of preferential alignment of particles in case of $A_z$=4 in direction $<10>$. This observations led us to conclude that the particles are always aligning in elastically soft direction to minimize its elastic energy. 

Matrix grain size and precipitate size evolutions are described in Fig.~\ref{anisotropy} (a) and (b) respectively. Coarsening kinetics of precipitates in anisotropic medium are much faster than the isotropic elastic medium. $A_z$=4 shows higher coarsening compared to  $A_z$=1/4  and that is why the alignment of the precipitates in the softest $<10>$ direction is more apparent here. But in matrix grain size $A_z$=4 and $A_z$=1/4 shows similar coarsening. Still elastic anisotropy increases average matrix grain size compared to the isotropic elasticity.

\subsection{Combined effect of $\delta$ and $A_z$}\label{combinesec}
\begin{figure}[htp]
\begin{center}
\includegraphics[width=\linewidth]{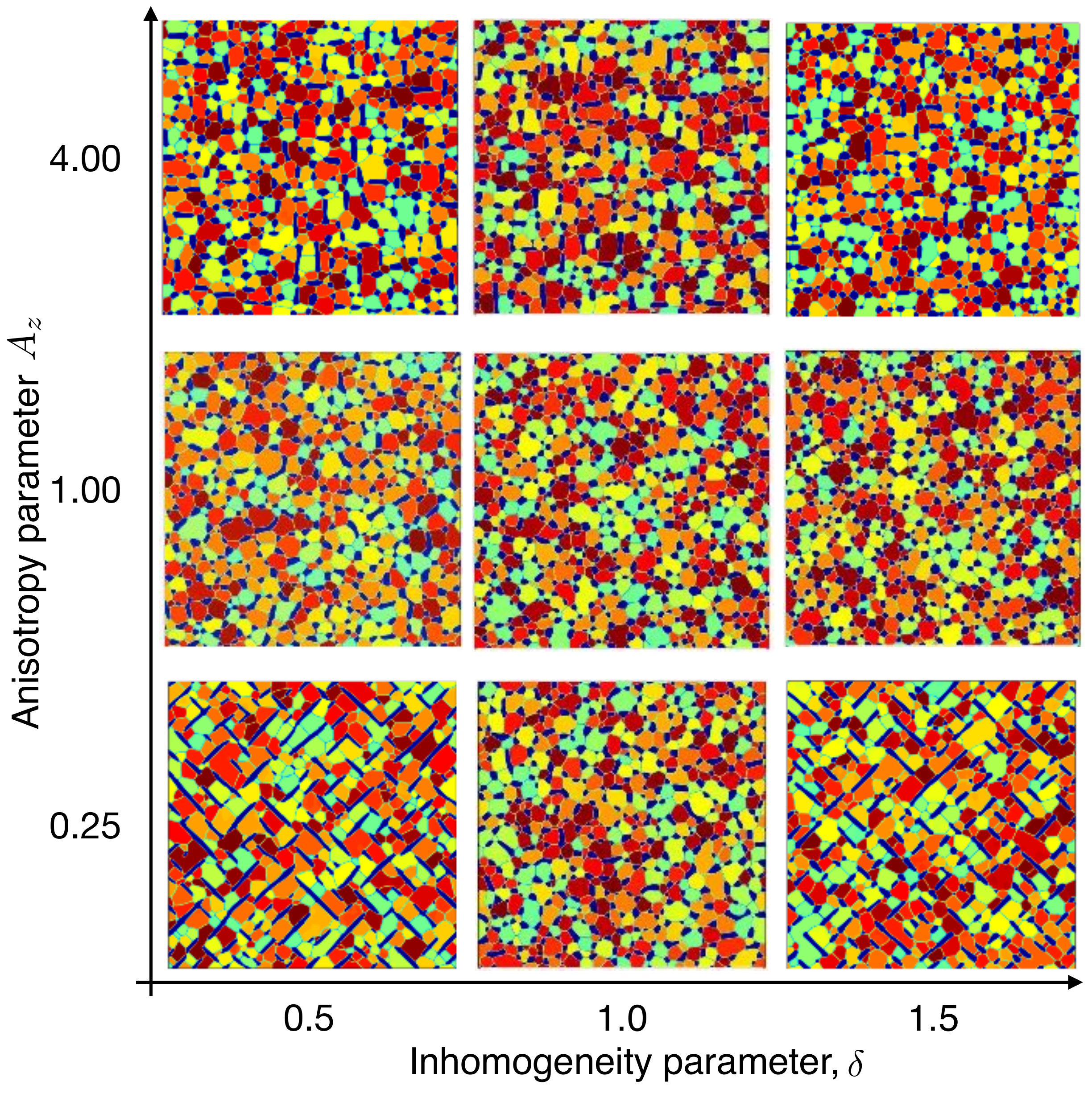}
\caption{Microstructural features in $\delta$-$A_z$ space. The elastic inhomogeneity parameter $\delta$ and elastic anisotropy parameter $A_z$ are varied in $x$ and $y$-directions respectively. For each considered microstructure, the strength of misfit ($\epsilon$) is 1.5$\%$.\label{combined-micro}}
\end{center}
\end{figure}
A real system will have elastic inhomogeneity and elastic anisotropy together. Therefore, it is important to understand the combined effect of these parameters. In Fig.~\ref{combined-micro}, we have shown the microstructural features by simultaneously varying values of elastic inhomogeneity ($\delta$) and elastic anisotropy($A_z$). Elastic anisotropy of 0.25 in case of $\delta$ equals 0.5 and 1.5 leads to needle shape microstructure preferentially oriented in $<11>$ direction as $<11>$ direction is the softest direction in case of $A_z<1$. Ardell \textit{et. al.} \cite{ARDELL19661295} have shown that the coherent misfit between the matrix and precipitate combining with elastic anisotropy in the crystal results in a gradual evolution from randomly aligned spherical precipitates to preferentially aligned needle shaped precipitates with longer coarsening times in Ni based super alloys. 

\begin{figure}[htp]
\begin{center}
\includegraphics[width=\linewidth]{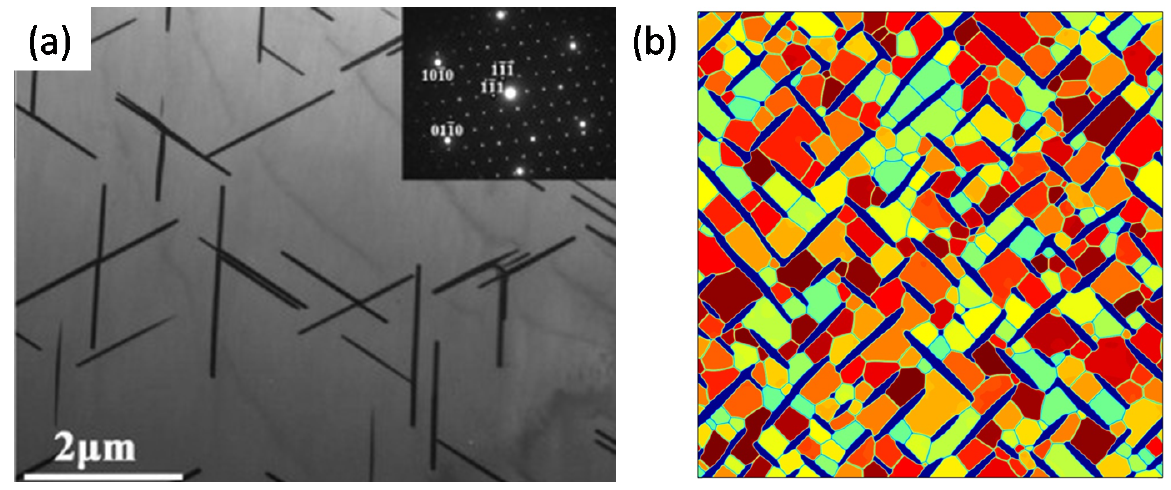}
\caption{Microstructural comparison between experimental observation and our phase field model. (a) TEM micrograph of needle shaped $\beta$ precipitates in magnesium based alloy (reprinted from  the works of Ting Li and co-workers\cite{Exp_mag_alloy} with permission from Elsevier\cite{Exp_mag_alloy}), 
(b) Our phase field model showing similar needle shaped precipitate on the grain boundaries $\delta$=0.5, $A_z$=0.25.\label{Exp_Comp}}
\end{center}
\end{figure}

Such needle shaped precipitates have been observed in many different systems experimentally. Figure \ref{Exp_Comp} (a) shows the TEM image of magnesium alloy after precipitation treated at 400$\degree$C. Here we can observe needle-shaped $\beta$ phase precipitates with length of around 3 $\mu$m. Most of these precipitates also occur on the grain boundaries and follows a orientation relationship with the matrix phase. Our results show such needle shaped precipitates can arise in polycrystalline matrix with elastic inhomogeneity and anisotropy \ref{Exp_Comp} (b).

\begin{figure}[htp]
\begin{center}
\includegraphics[width=0.9\linewidth]{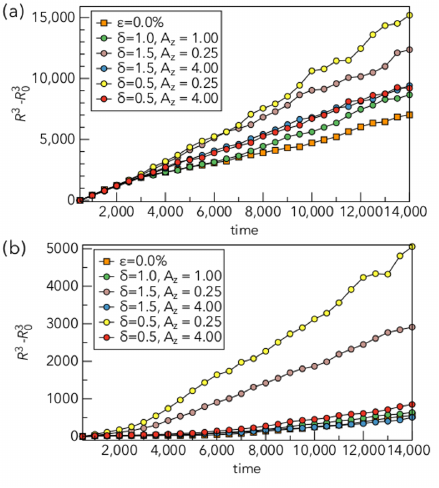}
\caption{Combined effect of elastic inhomogeneity, $\delta$ and elastic anisotropy, $A_z$ on (a) grain growth kinetics (b) precipitate kinetics\label{combined-kinetics}}
\end{center}
\end{figure}

Similar elongated micro-structures have also been observed in case of elastic anisotropy of 4.0 and $\delta$ 0.5 and 1.5. But here the precipitates are oriented in $<10>$ direction which is the softest direction in case of $A_z>1$. Without elastic inhomogeneity, preferential alignment of the precipitate in a particular direction has not been observed as in this case there is no difference elastic modulus wise between matrix and the precipitate phases.

Figure \ref{combined-kinetics} (a) and (b) show the combined effect of elastic inhomogeneity ($\delta$) and elastic anisotropy ($A_z$) on matrix grain and precipitate coarsening respectively. We have observe in sections \ref{deltasec},\ref{azsec}  that the elastic inhomogeneity and anisotropy both result in increase in precipitate size and corresponding increase in average grain size. Interestingly, preferentially oriented micro-structures in case of $A_z$=1/4 is most ineffective in inhibiting the matrix grain growth. On the other hand, systems with $A_z$=4.0 shows similar grain growth as the elastically isotropic precipitate. 

Previously, the effect of particle shape on Zener pinning has been studied by phase field modeling. \citep{chang2009effect} Their study reveals that second phase particle with higher aspect ratio such as particles with the needle shape morphology are more effective in retarding the grain growth. The total second phase particle surface area is a function of its shape and sphere has the smallest surface area per unit volume. Thus the sphere shape is least effective in Zener pinning and increase in aspect ratio increases the pinning effect. Our work shows such is not true in case of higher aspect ratio coherent precipitates. In earlier models, they have introduced the higher aspect ratio particles from the start and they have not included any elastic interaction. Also they do not include the effect of coarsening of second phase particles. In this work we have started with spherical particle which gradually changes to higher aspect ratio needle shape due to elastic interaction. We can say that our model predicts in this particular case the needle shaped precipitates do not help in retarding matrix grain growth through Zener pinning. The same elastic interaction which makes this needle shaped precipitate also increases the coarsening rate of the precipitate. Faster coarsening accompanying the needle shaped precipitate formation therefore renders the increased pinning of a needle shaped precipitate ineffective. 
 
\section{\label{sec:level4}Conclusions}
In this work, we have systematically studied the effect of elastic misfit strain, elastic inhomogeneity, and elastic anisotropy of the coherent precipitate on the Zener pinning of the matrix grains. The coarsening exponent of the temporal power law for coherent precipitate remain cubic as also been observed experimentally.\cite{lund2002effects} Increase in misfit strain, elastic inhomogeneity, and anisotropy increases the rate of precipitate coarsening which in turn also increases the grain growth in the matrix phase. Finally, the needle shaped precipitate formed due to elastic anisotropy does not help in Zener pinning due to increased coarsening in the precipitate from elastic interaction.

\textit{\textbf{Acknowledgement}} The author TC would like to thank DST for the financial support through Inspire Faculty Award (DST/INSPIRE/04/2017/000548) and also IIT Patna for the facilities.

\section{\label{sec:level6}References}
\bibliography{Phase_field}
\end{document}